\documentclass[english]{article}
\usepackage[T1]{fontenc}
\usepackage[latin9]{inputenc}
\usepackage{geometry}
\usepackage{array}
\usepackage{multirow}
\usepackage{amsmath}
\usepackage{amssymb}
\usepackage{stackrel}
\usepackage{graphicx}
\usepackage{setspace}
\PassOptionsToPackage{normalem}{ulem}
\usepackage{ulem}

\makeatletter

\providecommand{\tabularnewline}{\\}

\makeatother

\usepackage{babel}
\begin{document}
\begin{center}
{\LARGE{}Multi-scale information content measurement method based
on Shannon information}{\LARGE\par}
\par\end{center}

\begin{singlespace}
\begin{center}
{\small{}Zsolt Pocze}{\small\par}
\par\end{center}

\begin{center}
\textit{Volgyerdo Nonprofit Kft.}, Nagybakonak, HUN, info@volgyerdo.hu,
2023
\par\end{center}
\end{singlespace}
\begin{abstract}
In this paper, we present a new multi-scale information content calculation
method based on Shannon information (and Shannon entropy). The original
method described by Claude E. Shannon and based on the logarithm of
the probability of elements gives an upper limit to the information
content of discrete patterns, but in many cases (for example, in the
case of repeating patterns) it is inaccurate and does not approximate
the true information content of the pattern well enough. The new mathematical
method presented here provides a more accurate estimate of the (internal)
information content of any discrete pattern based on Shannon's original
function. The method is tested on different data sets and the results
are compared with the results of other methods like compression algorithms.
\end{abstract}

\section{Introduction}

Traditionally, Shannon's information theory \cite{SHANNON_MTC} has
been used to measure the information content of samples. Shannon information,
as defined by Claude E. Shannon, is the degree of uncertainty or surprise
associated with a given outcome in a set of possible outcomes. Shannon
entropy, which is the expected value of Shannon information, is used
to quantify the average information content of a discrete sample or
message. It serves as a basic concept in information theory and is
widely used in communication systems and data compression.

In some situations, such as repeated patterns, Shannon's original
information measurement method does not give accurate results enough
because it does not take into account the structure of the patterns,
it only looks at certain statistical characteristics of them. To solve
this problem, this paper presents a new multiscale information content
calculation method based on Shannon's original principles. By refining
the computational approach, our method offers a more accurate estimate
of the internal information content of discrete samples, regardless
of their nature.

There are several other methods for measuring the information content
of patterns, such as Kolmogorov complexity \cite{KOLMOGOROV}, randomness
\cite{LOVASZ_COMPLEXITY}, and compression complexity. The common
property of these methods that they are all suitable for determining
and understanding the information content of patterns with some accuracy,
and therefore provide a suitable comparison basis for checking newer
methods.

To verify the effectiveness of our new method, it is applied to various
data sets and compared with compression algorithms. The results show
that our proposed method based on Shannon information closely approximates
the results measured by other methods while taking a completely different
approach.\pagebreak{}

\section{{\large{}Patterns}}

In this study, we deal with the calculation of the internal quantitative
information content of discrete patterns. From the point of view of
the calculation of the information content, the nature of the object
of the measurement is irrelevant. The information content of events,
signals, system states, or data sequences can be calculated since
their models (with finite precision) can all be represented as discrete
patterns. By moving along a spatial pattern, we get a temporal pattern
and vice versa. Therefore, we do not distinguish between spatial and
temporal patterns. The basic markings should be as follows.

Denote$\mathcal{M}(R)$ the set of finite sequences that can be generated
from the set $R$:

\begin{equation}
\mathcal{M}(R)=\{X:\mathbb{N}^{+}\rightarrow R\}
\end{equation}

Let us call the finite sequence $X\in\mathcal{M}(R)$ a pattern:

\begin{equation}
X=[x_{1},...,x_{N}]
\end{equation}

Denote the length of the series $X$:

\begin{equation}
n(X)=N
\end{equation}

Denote the set of possible values of the series $X$:

\begin{equation}
R=R_{X}=\{r_{1},r_{2},...,r_{K}\}
\end{equation}

Let $f(x)$ denite the number of occurrences of $x\in R_{X}$ in the
series of $X$:

\begin{equation}
f(x)=\sum_{i=1}^{K}[r_{i}=x]
\end{equation}

Let the relative frequency of any $x\in R$ element of the pattern
$X$:

\begin{equation}
p(x)=f(x)/N
\end{equation}

Denote the concatenation of $X_{1}X_{2}...X_{K}$ patterns as:

\begin{equation}
X_{1}X_{2}...X_{K}=\stackrel[i=1]{K}{\Vert}X_{i}
\end{equation}
\pagebreak{}

\section{{\large{}Information content}}

The information content can be interpreted intuitively when only the
interpretable information content is examined \cite{FACTICITY_SELF_DESCRIPTIVE_INFO}.
In this study we examine the amount of the entire internal information
content without interpreting it or considering the context.

The information content of a pattern can be characterized by the improbability
of individual elements of the pattern (Shannon information \cite{SHANNON_MTC}),
the length of the most concise description of the pattern (Kolmogorov
complexity \cite{KOLMOGOROV}), or the degree of randomness of the
pattern \cite{LOVASZ_COMPLEXITY}.

A fundamental difference between Shannon's and Kolmogorov's viewpoints
is that Shannon considered only the probabilistic characteristics
of the random source of information that created the pattern, ignoring
the pattern itself. In contrast, Kolmogorov only focused on the pattern
itself \cite{SHANNON_KOLMOGOROV}. In their definition, Kolmogorov
and Chaitin called (inaccurately) random the pattern with the maximum
information content \cite{COMPLEXITY_ALGORITHMS}.

Information, complexity and randomness have such similar properties
that we can reasonably assume that they are essentially approaching
the same thing with different methods. It is sufficient to consider
that the Shannon information, Kolmogorov complexity and randomness
of a pattern consisting of identical elements are all minimal, while
in the case of a true random pattern all three values are maximal,
and they all assign the highest information value to data sets with
maximum entropy \cite{FACTICITY_SELF_DESCRIPTIVE_INFO}.

The concepts of entropy and information are often confused \cite{ENTROPY_USES_MISUSES},
so it is important to mention that entropy can also be understood
as such the average information content per element.

Approached intuitively, the amount of information is a function for
which the following conditions are met:
\begin{enumerate}
\item The information content of a pattern with zero length or consisting
of identical elements is zero.
\item The information content of the pattern consisting of repeating sections
is (almost) identical to the information content of the repeating
section.
\item A pattern and its reflection have the same information content.
\item The sum of the information content of patterns with disjoint value
sets is smaller than the information content of the concatenated pattern.
\item The information content of true random patterns is almost maximal.
\end{enumerate}
Let the information content be the function $I$ that assigns a non-negative
real number to any arbitrary pattern $X\in\mathcal{M}(R)$:

\begin{equation}
I:\mathcal{M}_{R}\rightarrow\mathbb{R}^{+}
\end{equation}
\pagebreak{}

In addition, the following conditions are met:
\begin{enumerate}
\item $I(X)=0\Leftrightarrow|R_{X}|<2$
\item $I(\stackrel[i=1]{K}{\Vert}X)=I(X)$
\item $I(\stackrel[i=1]{K}{\Vert}X_{i})=I(\stackrel[i=K]{1}{\Vert}X_{i})$
\item $|\stackrel[i=1]{K}{\cap}R_{X_{i}}|=\emptyset\Rightarrow I(\stackrel[i=1]{K}{\Vert}X_{i})>\stackrel[i=1]{K}{\sum}I(X_{i})$
\item $I(X)\leq I(X_{TR}),\forall X\in\mathcal{M}(R),|X|=|X_{TR}|$, where
$X_{TR}\in\mathcal{M}(R)$ is a real random pattern.
\end{enumerate}
Since any pattern can be described in non-decomposable binary form,
the unit of information content should be the bit.

It can be seen that for any pattern $X\in\mathcal{M}(R)$, if $N=n(X)$
and $K=|R|$, then the maximum information content of $X$ is:

\begin{equation}
I_{MAX}(X)=N\cdot log_{2}(K)
\end{equation}

That is, $I(X)\leq I_{MAX}(X)$ for any pattern $X\in\mathcal{M}(R)$.
In the case of a binary pattern, $I_{MAX}(X)=N$, the length of the
pattern, which means that a maximum of $N$ bits of information (decision)
is required to describe the pattern.

If the maximum information content is known, the relative information
content can be calculated:

\begin{equation}
I^{(rel)}(X)=I(X)/I_{MAX}(X)
\end{equation}

\section{{\large{}Shannon information}}

In theory, Kolmogorov complexity would provide a better approximation
of the information content of patterns, but it has been proven that
it cannot be calculated \cite{SHANNON_KOLMOGOROV}, in contrast to
Shannon information \cite{SHANNON_MTC}, which can be calculated efficiently,
but approximates the actual information content less well. Shannon
information calculates the information content of the pattern based
on the expected probability of occurrence (relative frequency) of
the elements of the pattern.

The Shannon information of an arbitrary pattern $X\in\mathcal{M}(R)$:

\begin{equation}
I_{S}(X)=\sum_{i=1}^{N}log_{2}(\frac{1}{p(x_{i})}))
\end{equation}

Since the relative frequency (expected occurrence) of the elements
of the pattern is only one statistical characteristic of the pattern
and does not take into account the order of the elements. That's why
the Shannon information often gives a very inaccurate estimate of
the information content. \pagebreak{}

The value of the Shannon information is the same for all patterns
of the same length whose elements have the same relative frequency.
If $X\in\mathcal{M}(R)$, $Y\in\mathcal{M}(Q)$ and $|R|=|Q|=K$ then
it holds that:

\begin{equation}
I_{S}(X)=I_{S}(Y),\:if\:\{p(r_{1}),p(r_{2}),...,(r_{K})\}=\{p(q_{1}),p(q_{2}),...,(q_{K})\}
\end{equation}

Shannon information ignores the structure of the patterns at different
scales, the laws encoded in them, and therefore overestimates the
information content of patterns consisting of repeating sections.

The problem can be illustrated with a simple example. Let's calculate
the Shannon entropy of the following three patterns:
\begin{enumerate}
\item $X_{A}:\:001101101010111001110010001001000100001000010000$
\item $X_{B}:\:101010101010101010101010101010101010101010101010$
\item $X_{C}:\:111111110000000011111111000000001111111100000000$
\end{enumerate}
In all three cases, the set of values is $R={0,1}$, the probability
of each element is $p(0)=0.5$ and $p(1)=0.5$, and the Shannon entropy
is $I_{S}(X)=\stackrel[i=1]{N}{\sum}log_{2}(\frac{1}{p(x_{i})})=16\:bit$,
although it is obvious that the information content of the data series
differs significantly. Due to its randomness, the information content
of data line $X_{A}$ is almost 16 bits, while the information content
of the other two data lines is much smaller, as they contain repeated
sections. In the $X_{B}$ data line, for example, the 2-bit section
$[10]$ is repeated, which means that its information content is closer
to 2 bits.

The problem is that in the example above, we are examining the datasets
at an elementary level, and our Shannon entropy function does not
take into account the larger-scale structure of the dataset, such
as the presence of repeating segments longer than 1 signal. Therefore,
it is obvious to develop methods that are based on the Shannon entropy,
but the data series are analyzed in the entire spectrum of the resolution,
in the entire frequency range, and thus provide a more accurate approximation
of the information content of the data series. Countless such solutions
have already been published, which can be read, for example, in the
articles \cite{MULTISCALE_THEORY} and \cite{MULTISCALE_ENTROPY_REVIEW}.
This article presents an additional method.

\section{{\large{}SSM information}}

\subsection{Shannon information spectrum}

Let the pattern $X$ be partitioned by sections of length $r$ if
$m=[N/r]$:

\begin{equation}
X^{(r)}=[x_{1}...x_{r},x_{r+1}...x_{2r},\:...,\:x_{(m-1)\cdot r+1}...x_{m\cdot r}]
\end{equation}

Let the following series denoted as Shannon information spectrum (SP)
of pattern $X$:

\begin{equation}
I_{SP}^{(r)}(X)=I_{S}(X^{(r)}),\:r=1,...,[N/2]
\end{equation}

From the sequences $X^{(r)}$, we omit (truncated) partitions shorter
than $r$, those that are shorter than $r$. In the cases $r>[N/2]$,
$I_{SP}(X^{(r)})=0$, so these are also omitted from the spectrum.
\begin{center}
\includegraphics[width=0.8\paperwidth]{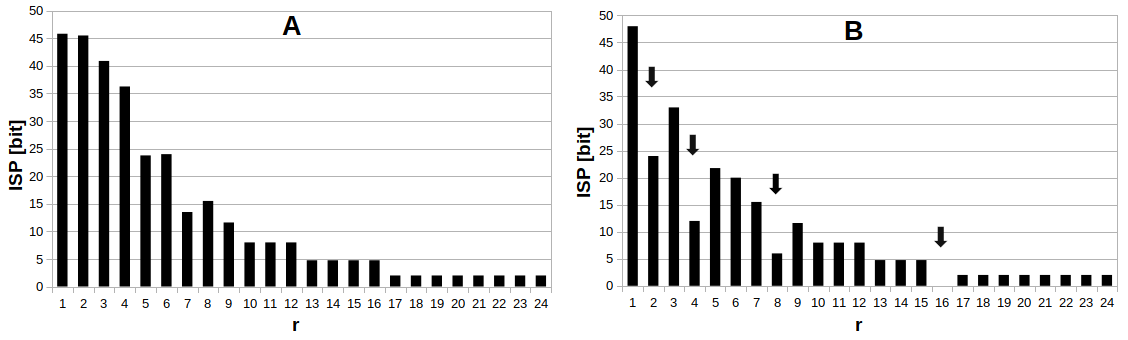}
\par\end{center}

\begin{center}
\textbf{\textit{\footnotesize{}Figure 1.}}\textit{\footnotesize{}
Diagram $A$ shows the Shannon information spectrum of the random
pattern $X_{A}$, and diagram $B$ shows the repeating pattern$X_{C}$
(Appendix I). It can be seen that in case $B$, a lower value appears
at certain frequencies.}{\footnotesize\par}
\par\end{center}

\subsection{Maximal Shannon information spectrum}

The Shannon information spectrum will be maximum in the case of random
data sets. Let the following formula denoted as maximum Shannon information
spectrum (SMS):

\begin{equation}
I_{SMS}^{(r)}(X)=m\cdot log_{2}(min(K^{r},m)),\:r=1,...,[N/2]
\end{equation}

$I_{SMS}^{(r)}(X)$ is a supremum for all information spectrum having
the same value set and pattern length. If $K^{r}<m$, then in the
case of random patterns, the value set of the partitioning contains
most likely all possible partitions, so the information content is
approximately $m\cdot log_{2}(K^{r})$. If $K^{r}>m$, then the partitioning
cannot contain all possible partitions, each partition will most likely
be unique, so the information content will be $m\cdot log_{2}(m)$.
If $r$ is small enough, then the series $X^{(r)}$ most likely contains
all possible partitions, therefore by random data sets the measured
amount of information will approximately equal the maximum possible
information content of the pattern, i.e. if $r$ is small, then $I_{SPM}^{(r)}(X^{(r)})\approx I_{MAX}(X^{(r)})=N\cdot log_{2}(n)$.
\begin{center}
\includegraphics[width=0.5\columnwidth]{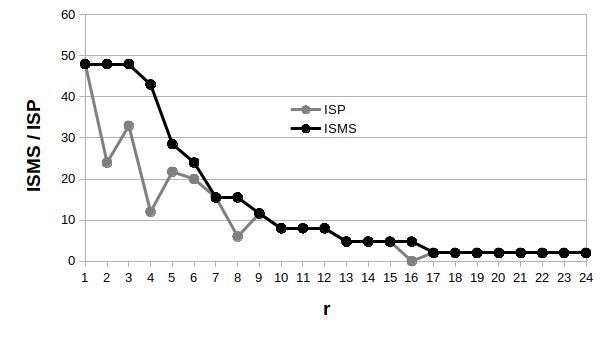}
\par\end{center}

\begin{center}
\textbf{\textit{\footnotesize{}Figure 2. }}\textit{\footnotesize{}Comparison
of maximum Shannon information spectrum (ISMS) and Shannon information
spectrum (ISP) of the repeating pattern $X_{C}$.}{\footnotesize\par}
\par\end{center}

\subsection{Shannon normalized information spectrum}

If we are interested in how much the information content seems to
be relative to the maximum value in each resolution, we can normalize
the spectrum with the maximum spectrum to the range $[0-N\cdot log_{2}(n)]$.
Let the following sequence denoted as Shannon normalized information
spectrum (SNS):

\begin{equation}
I_{SNS}^{(r)}(X)=\begin{cases}
\frac{I_{SP}^{(r)}(X)}{I_{SMS}^{(r)}(X)}\cdot I_{MAX}(X), & if\:|R_{X^{(r)}}|>1\\
r\cdot\frac{I_{SP}^{(1)}(X)}{N}, & if\:|R_{X^{(r)}}|=1
\end{cases}\quad where\:r=1,...,[N/2]
\end{equation}

If the value set of the partitioning has only one element, i.e. $|R_{X^{(r)}}|=1$,
the normalized value would be $0$. In this case the information content
should be the information content of the repeating partition, and
the average Shannon entropy of an element of the elementary resolution
is multiplied by the length of the partition: $r\cdot\frac{I_{SP}^{(1)}(X)}{N}$.
\begin{center}
\includegraphics[width=0.8\paperwidth]{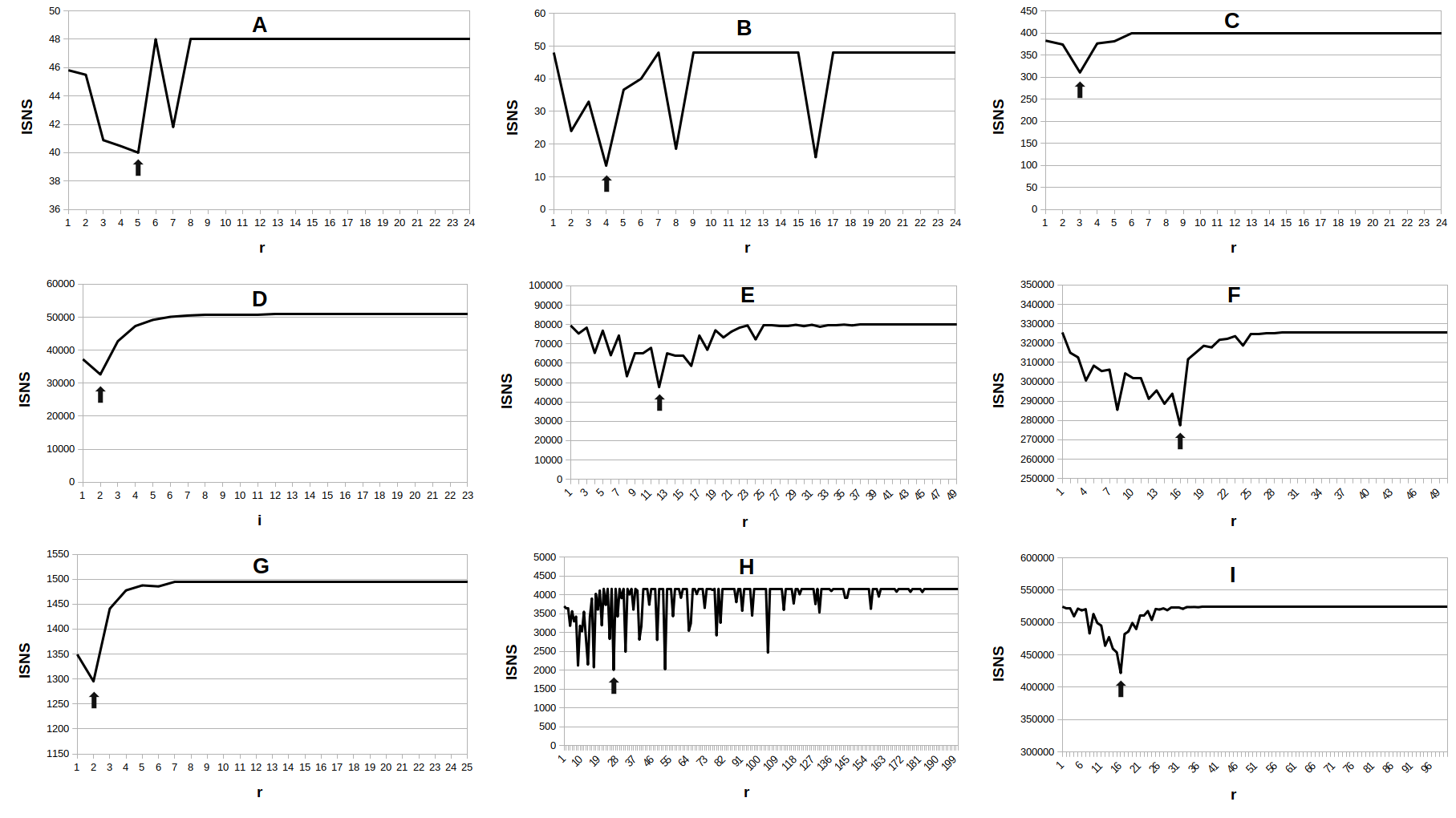}
\par\end{center}

\begin{center}
\textbf{\textit{\footnotesize{}Figure 3. }}\textit{\footnotesize{}Comparison
of Shannon normalized information spectrum (S of patterns from very
different sources. The vertical axis represents the amounts of bits
of information measured at the given resolution. You can see how different
the spectrum of the different patterns is, but in most cases there
is a resolution where the information content shows a definite minimum.
The minima are marked with an arrow. A: random binary pattern, B:
binary pattern with repeating sections, C: DNA section, D: English
text, E: ECG signal, F: audio recording containing speech, G: evolution
of the number of sunspots between 1700-2021, H: seismogram , I: Lena's
photo.}{\footnotesize\par}
\par\end{center}

The figures show that different types of patterns have very different
and characteristic spectra. This suggests that the type or source
of the pattern may be inferred from the nature of the spectrum, but
we do not deal with this in this study.

\subsection{SSM information}

We know that the Shannon information gives an upper estimate in all
cases, so we get the most accurate approximation of the information
content from the normalized spectrum if we take the minimum. Let the
information content calculated from the normalized spectrum denoted
as Shannon spectrum minimum information (SSM information):

\begin{equation}
I_{SSM}(X)=\stackrel[i=1]{[N/2]}{min}(I_{SNS}^{(i)}(X))
\end{equation}

Shannon information, SSM information and compression complexity of
different patterns (Appendix I) in bits:
\begin{center}
\begin{tabular}{|c|l|c|c|c|c|c|c|}
\hline 
Pattern & Source & $I_{MAX}(X)$ & $I_{S}(X)$ & $I_{SSM}(X)$ & $I_{ZIP}(X)$ & $I_{7Z}(X)$ & $I_{ZPAQ}(X)$\tabularnewline
\hline 
\hline 
X$_{A}$ & {\scriptsize{}Random binary pattern.} & 48 & 46 & 40 &  &  & \tabularnewline
\hline 
X$_{B}$ & {\scriptsize{}Repeating binary pattern.} & 48 & 48 & 2 &  &  & \tabularnewline
\hline 
X$_{C}$ & {\scriptsize{}Repeating binary pattern.} & 48 & 48 & 13 &  &  & \tabularnewline
\hline 
X$_{D}$ & {\scriptsize{}Repeating text.} & 362 & 343 & 58 &  &  & \tabularnewline
\hline 
X$_{E}$ & {\scriptsize{}Duplicate text with one character error.} & 374 & 347 & 116 &  &  & \tabularnewline
\hline 
X$_{F}$ & {\scriptsize{}Random DNA pattern.} & 471 & 422 & 409 &  &  & \tabularnewline
\hline 
X$_{G}$ & {\scriptsize{}DNA segment of COVID virus.} & 471 & 405 & 388 &  &  & \tabularnewline
\hline 
X$_{H}$ & {\scriptsize{}Random string (0-9, a-z, A-Z).} & 1209 & 1174 & 1174 &  &  & \tabularnewline
\hline 
X$_{I}$ & {\scriptsize{}English text (James Herriot's Cat Stories).} & 1104 & 971 & 971 &  &  & \tabularnewline
\hline 
X$_{J}$ & {\scriptsize{}Solar activity between 1700-2021 (A-Z).} & 1495 & 1349 & 1295 &  &  & \tabularnewline
\hline 
X$_{K}$ & {\scriptsize{}Isaac Asimov: True love.} & 50901 & 37266 & 32649 & 30904 & 29968 & 25248\tabularnewline
\hline 
X$_{L}$ & {\scriptsize{}Binary ECG signal.} & 80000 & 79491 & 47646 & 52320 & 41032 & 36968\tabularnewline
\hline 
X$_{M}$ & {\scriptsize{}Binary seismic data.} & 313664 & 312320 & 171546 & 83920 & 66064 & 45824\tabularnewline
\hline 
X$_{N}$ & {\scriptsize{}Speech recording.} & 325472 & 325342 & 277489 & 286760 & 257856 & 251408\tabularnewline
\hline 
X$_{O}$ & {\scriptsize{}Lena.} & 524288 & 524216 & 422085 & 443096 & 371360 & 337408\tabularnewline
\hline 
\end{tabular}
\par\end{center}

\begin{center}
\textbf{\textit{\footnotesize{}Table 1.}}\textit{\footnotesize{} Comparison
of SSM information and compression complexity of different patterns.}{\footnotesize\par}
\par\end{center}

Relative Shannon information, SSM information, and compression complexity
of different patterns (Appendix I) compared to maximum information:
\begin{center}
\begin{tabular}{|c|l|c|c|c|c|c|}
\hline 
Pattern & Source & $I_{S}^{(rel)}(X)$ \% & $I_{SSM}^{(rel)}(X)$ \% & $I_{ZIP}^{(rel)}(X)$ \% & $I_{7Z}^{(rel)}(X)$ \% & $I^{(rel)}{}_{ZPAQ}(X)$ \%\tabularnewline
\hline 
\hline 
X$_{K}$ & {\scriptsize{}Isaac Asimov: True love.} & 73 & 64 & 61 & 59 & 50\tabularnewline
\hline 
X$_{L}$ & {\scriptsize{}Binary ECG signal.} & 99 & 60 & 65 & 51 & 46\tabularnewline
\hline 
X$_{M}$ & {\scriptsize{}Binary seismic data.} & 100 & 55 & 27 & 21 & 15\tabularnewline
\hline 
X$_{N}$ & {\scriptsize{}Speech recording.} & 100 & 85 & 88 & 79 & 77\tabularnewline
\hline 
X$_{O}$ & {\scriptsize{}Lena.} & 100 & 81 & 85 & 71 & 64\tabularnewline
\hline 
\end{tabular}
\par\end{center}

\begin{center}
\textbf{\textit{\footnotesize{}Table 2.}}\textit{\footnotesize{} Comparison
of relative SSM information and relative compression complexity of
different patterns.}{\footnotesize\par}
\par\end{center}

It can be seen from the table that the SSM information gives similar
results as the compression algorithms. In general, it is true that
the more computationally demanding a compression or information measurement
procedure is, the closer it is to Kolmogorov complexity. In the examined
examples, the results of SSM information are usually located between
the results of ZIP and 7Z, so the computational complexity of SSM
information must be similar to the computational complexity of ZIP
and 7Z.
\begin{center}
\includegraphics[width=0.4\paperwidth]{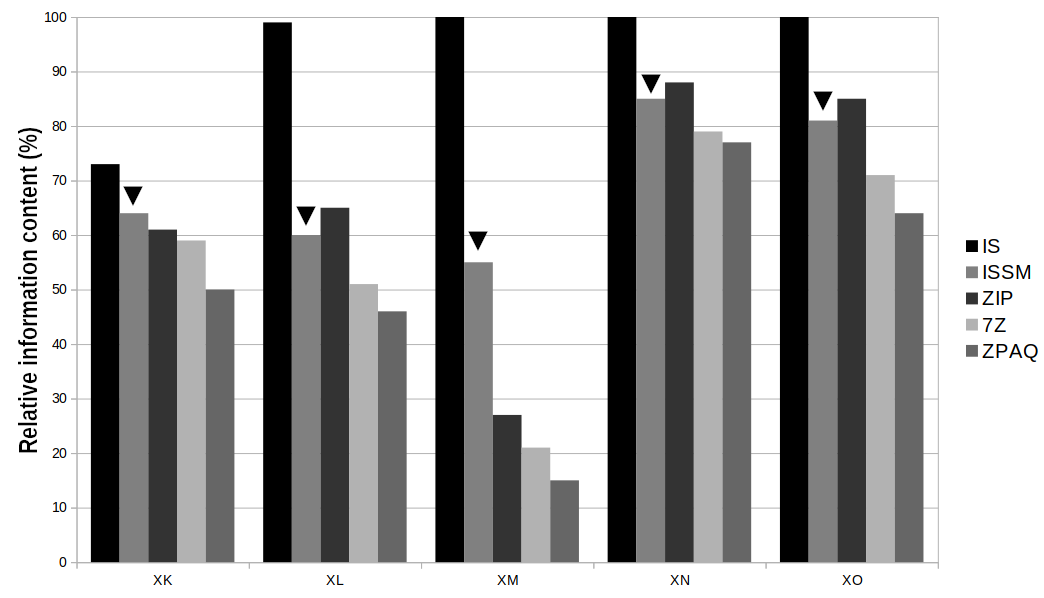}
\par\end{center}

\begin{center}
\textbf{\textit{\footnotesize{}Figure 4.}}\textit{\footnotesize{}
Comparison of the results of different information measurement methods.}{\footnotesize\par}
\par\end{center}

\begin{center}
\includegraphics[width=0.3\paperwidth]{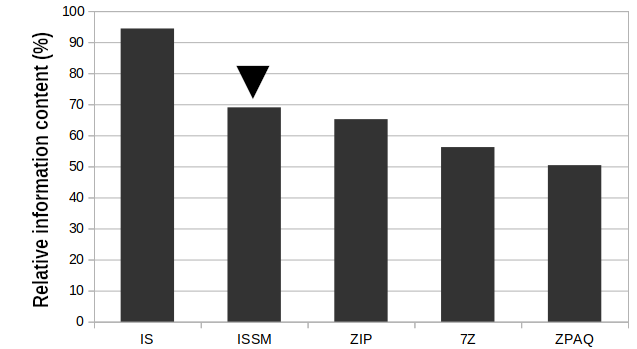}
\par\end{center}

\begin{center}
\textbf{\textit{\footnotesize{}Figure 5. }}\textit{\footnotesize{}Comparison
of the average results of different information measurement methods.}{\footnotesize\par}
\par\end{center}

\subsection{Comparison with computational complexity}

If we do not know the signal set of the signal sequence, the first
step is to determine the number of signals occurring in the signal
sequence, which has an asymptotic complexity of $\mathcal{O\mathrm{(}}N\cdot logN)$.

Determining the Shannon information consists of two steps. In the
first step, we determine the frequency of signals, which has a complexity
of $\mathcal{O\mathrm{(}}N)$, and in the second step, we sum up the
entropy of each signal, so the total complexity of the Shannon information
is $\mathcal{O\mathrm{(}}N\cdot logN)+\mathcal{O\mathrm{(}}N)=\mathcal{O\mathrm{(}}N\cdot logN)$.

For the ZIP, 7Z and ZPAQ algorithms used to calculate the compression
complexity, the complexity is usually between $\mathcal{O\mathrm{(}}N)$
and $\mathcal{O\mathrm{(}}N\cdot logN)$, but for ZPAQ may be greater
\cite{FAST_BWT} \cite{AUTOCORRELATED_CONTEXT_MODELS} \cite{REPETITIVE_COMPRESSION}.

In the case of SSM information, the first step is also to determine
the frequency of signals, which has a complexity of $\mathcal{O\mathrm{(}}N)$.
In the second step, the Shannon information spectrum is calculated
$\mathcal{O\mathrm{(}}N)+\mathcal{O\mathrm{(}}N/2)+\mathcal{O\mathrm{(}}N/3)+...+\mathcal{O\mathrm{(}}2)=\mathcal{O\mathrm{(}}N\cdot logN)$
complexity, finally the minimum of the spectrum can be determined
$\mathcal{O\mathrm{(}}N)$ with complexity. The complexity of calculating
the SSM information in the worst case is $\mathcal{O\mathrm{(}}I_{SSM}(X))=\mathcal{O\mathrm{(}}N\cdot logN)+\mathcal{O\mathrm{(}}N)+\mathcal{O\mathrm{(}}N\cdot logN)+\mathcal{O\mathrm{(}}N)=\mathcal{O\mathrm{(}}N\cdot logN)$
, which is identical to compression algorithms.\pagebreak{}

\subsection{Known issues}

All methods of calculating the amount of information have inaccuracies.
One of the problems with SSM information is that if the repetition
in a repeating pattern is not perfect, the value of the SSM information
is larger than expected, as shown in the example below.
\begin{center}
\begin{tabular}{|c|c|}
\hline 
$X$ & $I_{SSM}(X)${[}bit{]}\tabularnewline
\hline 
\hline 
\texttt{123456789 123456789 123456789} & 29\tabularnewline
\hline 
\texttt{\uline{2}}\texttt{23456789 123456789 123456789} & 50\tabularnewline
\hline 
\end{tabular}
\par\end{center}

\begin{center}
\textbf{\textit{\footnotesize{}Table 3.}}\textit{\footnotesize{} One
element change can cause a notable difference in SSM information.}{\footnotesize\par}
\par\end{center}

\section{Conclusion}

It can be shown SSM information can determine the information content
of the patterns with an accuracy comparable to the compression algorithms,
but at the same time it is simple. In addition information spectrum
presented here provides a useful visual tool for studying the information
structure of patterns in the frequency domain.

\bibliographystyle{plain}
\nocite{*}
\bibliography{Multi-scale_information_content_measurement_method_based_on_Shannon_information}

\begin{center}
\texttt{\textbf{\Large{}\newpage{}Appendix}}{\Large\par}
\par\end{center}

\begin{center}
\textbf{I. Example patterns}
\par\end{center}

\begin{center}
\begin{tabular}{|c|l|c|l|}
\hline 
\textbf{Notation} & \textbf{A pattern or a detail of the pattern} & \textbf{Length} & \textbf{Explanation}\tabularnewline
\hline 
\hline 
$X_{A}$ & \texttt{\scriptsize{}001101101010111001110010001001000100001000010000} & {\scriptsize{}48 bit} & {\scriptsize{}Random binary pattern.}\tabularnewline
\hline 
$X_{B}$ & \texttt{\scriptsize{}101010101010101010101010101010101010101010101010} & {\scriptsize{}48 bit} & {\scriptsize{}Repeating binary pattern.}\tabularnewline
\hline 
$X_{C}$ & \texttt{\scriptsize{}111111110000000011111111000000001111111100000000} & {\scriptsize{}48 bit} & {\scriptsize{}Repeating binary pattern.}\tabularnewline
\hline 
\multirow{2}{*}{$X_{D}$} & {\scriptsize{}The sky is blue. The sky is blue. The sky is blue.} & \multirow{2}{*}{{\scriptsize{}101 characters}} & \multirow{2}{*}{{\scriptsize{}Repeating text.}}\tabularnewline
 & {\scriptsize{}The sky is blue. The sky is blue. The sky is blue.} &  & \tabularnewline
\hline 
\multirow{2}{*}{$X_{E}$} & {\scriptsize{}The sky is blue. The sky is blue. The sky is blue.} & \multirow{2}{*}{{\scriptsize{}101 characters}} & \multirow{2}{*}{{\scriptsize{}Duplicate text with one character error.}}\tabularnewline
 & {\scriptsize{}The sky is blue. The sky is glue. The sky is blue.} &  & \tabularnewline
\hline 
\multirow{4}{*}{$X_{F}$} & \texttt{\scriptsize{}cagtttctagctatattagcgggcacgactccactgcgcctatgcggaag} & \multirow{4}{*}{{\scriptsize{}200 characters}} & \multirow{4}{*}{{\scriptsize{}Random DNA pattern.}}\tabularnewline
 & \texttt{\scriptsize{}cttgatcaaattttgaccagatcttaggtaacctgaacaagtcagttcgt} &  & \tabularnewline
 & \texttt{\scriptsize{}aggcgtcgattggccgacgggtgcgaagaaaaaagtgatcgttgtccaac} &  & \tabularnewline
 & \texttt{\scriptsize{}atctctagtacccaccgttgtgatgtacgttatacggacacgagcatatt} &  & \tabularnewline
\hline 
\multirow{4}{*}{$X_{G}$} & \texttt{\scriptsize{}cggcagtgaggacaatcagacaactactattcaaacaattgttgaggttc} & \multirow{4}{*}{{\scriptsize{}200 characters}} & \multirow{4}{*}{{\scriptsize{}DNA segment of COVID virus.}}\tabularnewline
 & \texttt{\scriptsize{}aacctcaattagagatggaacttacaccagttgttcagactattgaagtg} &  & \tabularnewline
 & \texttt{\scriptsize{}aatagttttagtggttatttaaaacttactgacaatgtatacattaaaaa} &  & \tabularnewline
 & \texttt{\scriptsize{}tgcagacattgtggaagaagctaaaaaggtaaaaccaacagtggttgtta} &  & \tabularnewline
\hline 
\multirow{4}{*}{$X_{H}$} & \texttt{\scriptsize{}EK8Pi5sv2npTfzoaMNp87QtT5kbIUQkTJzHwICCstSmg4aksHT} & \multirow{4}{*}{{\scriptsize{}200 characters}} & \multirow{4}{*}{{\scriptsize{}Random string (0-9, a-z, A-Z).}}\tabularnewline
 & \texttt{\scriptsize{}MwztgHFg3j8AoIobN3FycCLidGeyROiNyG5itB9kxyez1LZjFF} &  & \tabularnewline
 & \texttt{\scriptsize{}HIBjipE7hidZyiJmilXM0mwnxzlzWSfQ0xP1OuFpWosMwS1cjY} &  & \tabularnewline
 & \texttt{\scriptsize{}t4nyv4ONx1FceWkAf8SdvDGZVzeVzq2EmOqRF6Im2iudcYRswj} &  & \tabularnewline
\hline 
\multirow{5}{*}{$X_{I}$} & \texttt{\scriptsize{}I think it was the beginning of Mrs. Bond's } & \multirow{5}{*}{{\scriptsize{}221 characters}} & \multirow{5}{*}{{\scriptsize{}English text (James Herriot's Cat Stories)}}\tabularnewline
 & \texttt{\scriptsize{}unquestioning faith in me when she saw me } &  & \tabularnewline
 & \texttt{\scriptsize{}quickly enveloping the cat till all you could } &  & \tabularnewline
 & \texttt{\scriptsize{}see of him was a small black and white head } &  & \tabularnewline
 & \texttt{\scriptsize{}protruding from an immovable cocoon of cloth.} &  & \tabularnewline
\hline 
$X_{J}$ & \multirow{1}{*}{\texttt{\scriptsize{}ABCDFIEDBBAAAABEHJJGEEDBDGMSPLHFBACFKMRPLGDCA{[}...{]}}} & {\scriptsize{}321 characters} & {\scriptsize{}Solar activity between 1700-2021 (A-Z).}\tabularnewline
\hline 
\multirow{3}{*}{$X_{K}$} & \texttt{\scriptsize{}My name is Joe. That is what my colleague, } & \multirow{3}{*}{{\scriptsize{}8391 characters}} & \multirow{3}{*}{{\scriptsize{}Isaac Asimov: True love.}}\tabularnewline
 & \texttt{\scriptsize{}Milton Davidson, calls me. He is a programmer
and} &  & \tabularnewline
 & \texttt{\scriptsize{}I am a computer program. {[}...{]}} &  & \tabularnewline
\hline 
$X_{L}$ & \texttt{\scriptsize{}1011000100110011101110111011001100110011{[}...{]}} & {\scriptsize{}80000 bit} & {\scriptsize{}Binary ECG signal \cite{PHYSIOLOGIC_SIGNALS}.}\tabularnewline
\hline 
$X_{M}$ & \texttt{\scriptsize{}110000101000000011000010100000001100001010000{[}...{]}} & {\scriptsize{}313664 bit} & {\scriptsize{}Binary seismic data.}\tabularnewline
\hline 
$X_{N}$ & \texttt{\scriptsize{}0101001001001001010001100100011011100100{[}...{]}} & {\scriptsize{}325472 bit} & {\scriptsize{}Speech recording.}\tabularnewline
\hline 
$X_{O}$ & \texttt{\scriptsize{}1010001010100001101000001010001010100011{[}...{]}} & {\scriptsize{}524288 bit} & {\scriptsize{}Lena (256x256 pixel, grayscale).}\tabularnewline
\hline 
\end{tabular}
\par\end{center}
\end{document}